\documentclass[twocolumn]{emulateapj} 

\newcommand{\vek}[1]{\mbox{\boldmath$#1$}}

\begin{document}

\title{Spontaneous current-layer fragmentation and cascading
  reconnection in solar flares: II. Relation to observations}

\author{Miroslav B\'{a}rta\altaffilmark{1,2,3}, 
J\"{o}rg B\"{u}chner\altaffilmark{1}, 
Marian Karlick\'{y}\altaffilmark{2}, and
Pavel Kotr\v{c}\altaffilmark{2}}
\affil{$^1$Max Planck Institute for Solar System Research,
D-37191 Katlenburg-Lindau, Germany\\
$^2$Astronomical Institute of the Academy of Sciences of the Czech
Republic, CZ-25165 Ond\v{r}ejov, Czech Republic\\
$^3$ Observatory Vla\v{s}im, CZ-25801 Vla\v{s}im, Czech Republic}
\email{barta@mps.mpg.de}


\begin{abstract}
In the paper by B\'arta et al. (ApJ, 2010) the authors addressed some open 
questions of the CSHKP scenario of solar flares by means of
high-resolution MHD simulations. They focused, in particular, 
on the problem of energy transfer from large to small
scales in decaying flare current sheet (CS).
Their calculations suggest, that magnetic flux-ropes (plasmoids) are formed 
in full range of scales by
a cascade of tearing and coalescence processes. Consequently, the
initially thick current layer becomes highly fragmented.
Thus, the tearing and coalescence cascade can cause
an effective energy transfer across the scales. 
In the current paper we investigate whether this mechanism
actually applies in solar flares. We extend the
MHD simulation by deriving model-specific features that can
be looked for in observations. The results of the
underlying MHD model showed that the plasmoid cascade creates a specific 
hierarchical distribution of non-ideal/acceleration
regions embedded in the CS. We therefore focus on the features 
associated with the fluxes of 
energetic particles, in particular on the structure and dynamics of emission
regions in flare ribbons. We assume that the structure and dynamics of
diffusion regions embedded in the CS imprint themselves into structure and 
dynamics of flare-ribbon kernels by means of magnetic-field mapping. 
Using the results of the underlying MHD simulation we derive the expected 
structure of ribbon emission and we extract selected statistical 
properties of the modelled bright kernels. 
Comparing the predicted emission and its properties with the observed 
ones we obtain a good agreement of the two.

\end{abstract}

 

\section{Introduction}
\label{sect:intro}

For many years the common picture of solar eruptions and flares is
based on the CSHKP model \citep[see][and references therein]{Magara+:1996}.
It involves reconnection in a global vertical flare current layer formed behind
ejected flux-rope/filament \citep[][and references therein]{Lin+Forbes:2000}.
The standard model is in good agreement with observed large-scale dynamics
of eruptive events. Several question, however,  remain open.
In particular, it is not clear, by which mechanism the energy accumulated
in relatively large-scale ($\approx 1000$~km) 
structures of magnetic field associated with
the flare CS is transfered towards the
small dissipation scales. Indeed, the magnetic diffusion in the almost
collisionless solar coronal plasma is an essentially kinetic process
at scales of the order $d_i=c/\omega_{pi}$ \citep{Buchner:2006}. 
Another open question is the relation between the well-organised
dynamics of solar eruptions observed 
at large scales and the HXR and radio signatures of fragmented energy
release \citep{Aschwanden:2002, Karlicky+:2000}. 
And finally, the CSHKP model
has been put in question by many authors \citep[e.g.][]{Vlahos:2007}, 
in particular because 
its apparent incapability to explain the large fluxes of accelerated
particles inferred from HXR observations.

Recently -- in order to address the issue of energy transfer across the broad
range of MHD scales -- \citet{Barta+:2010b} [in further text referred as
\textit{Paper~I}] investigated magnetic reconnection in an extended current
layer by means of high-resolution MHD simulation. 
Inspired by the ``\textit{fractal reconnection}'' conjecture of 
\citet{Shibata+Tanuma:2001} the authors of Paper~I performed a 2.5D
numerical simulation covering
large range of MHD scales. They conjectured, that -- in analogy with
the vortex-tube cascade in fluid dynamics -- a cascade of magnetic flux-tubes
from large to small scale can provide the mechanism for energy
transfer across the scales. Two mechanisms of fragmentation
were identified in Paper~I: (1) The tearing cascade -- in line with
concept of \textit{fractal reconnection} by \citet{Shibata+Tanuma:2001},
developed recently into the theory of \textit{chain plasmoid instability} by
\citet{Loureiro+:2007} and \citet{Uzdensky+:2010}, supported further by 
numerical MHD simulations by \citet{Bhattacharjee+:2009},
\citet{Samtaney+:2009}, and \citet{Huang+Bhattacharjee:2010} 
and, (2) ambient-field driven coalescence of flux-ropes/plasmoids
leading to formation
of transversal current sheets subjected further to the same chain of processes
of cascading
fragmentation. Extrapolation of the results leads to the picture in which
the plasmoid cascade continues down to the kinetic scale where actual
magnetic dissipation and particle acceleration occurs, 
most likely via the kinetic coalescence of plasmoids
\citep{Drake+:2005,Karlicky+:2010}. 
Nevertheless, other processes can appear in the range between MHD and
plasma kinetic scales, e.g., the Hall type of reconnection -- see
recent simulations by \citet{Shepherd+Cassak:2010} and \citet{Huang+:2010}.

As it has been shown in the Paper~I, 
cascading reconnection forms multiple 
non-ideal regions (at the resolution limit -- see discussion in
Paper~I) hierarchically distributed in the fragmenting current
layer. As a result, the open questions of energy
transfer, fragmented vs. organised flare energy release, and particle
acceleration seem to be closely related to each other in the presence
of cascading reconnection.

The question arises, whether this mechanism found by simulations
is relevant for actual solar flares. We address
this topic in our present paper. For this sake we derive critical
signatures specific to the model developed in Paper~I, that
allow comparison with observations. 
It is currently impossible to observe small-scale magnetic
structures, whose formation in the fragmented flare current layer is
predicted by the cascading-reconnection model, directly.
Almost all information on the impulsive phase of flares comes
from the radiation emitted by accelerated particles.
Therefore we concentrate on such predictions of the MHD model that are 
connected with specific distribution and dynamics of acceleration
regions in the fragmenting flare current layer.

In particular, by magnetic mapping of non-ideal regions embedded in
the current sheet to the photosphere we derive
expected consequences of the distribution and dynamics of acceleration
regions for flare ribbons. Using the results of underlying MHD model presented
in Paper~I we qualitatively derive the structure and dynamics of
expected ribbon emission. We compare the modelled ribbon structure
with observations directly and by means of its statistical
properties used by \citet{Nishizuka+:2009}.

The paper is organised as follows: First, in
Section~\ref{subsect:modmhd}, we briefly summarise the results of 
the underlying high-resolution MHD simulation of cascading reconnection. 
Then, in Section~\ref{subsect:modext} we present the extension of
underlying MHD model utilised to produce  model-specific 
results/consequences in the form able to be directly compared 
with observations.  
In particular, we study the distribution and dynamics of
dissipative/acceleration regions, represented by the embedded X-points,
and their magnetic mapping to the chromosphere/photosphere of the
Sun. For obtained dynamics of acceleration regions and their
magnetic foot-points we qualitatively model the expected emission
of flare ribbons (Section~\ref{sect:results}). 
We compare the emission structure of modelled and
observed ribbons. Furthermore, we extract
selected statistical properties of emission kernels embedded in
modelled ribbons and compare them with the results found by analysis
of actual ribbon observations \citep{Nishizuka+:2009}. Finally, 
in Section~\ref{sect:conclusions} we discuss our results with the
intention to evaluate the relevance of cascading
reconnection/fragmentation processes presented in Paper~I for actual
solar flares.


\section{Model}
\label{sect:model}

In this section we describe the procedures allowing to simulate observable
signatures specific to cascading reconnection. For clarity we first
briefly summarise the main features of current-layer fragmentation
model (Paper~I).

\subsection{Basic MHD model}
\label{subsect:modmhd}

In Paper~I the authors studied cascading reconnection and energy
transfer from accumulation (large) to dissipation (small) scales. 
They used a high-resolution MHD model with a non-ideal, resistive term 
depending on the current-carrier drift velocity 

$v_{\rm D}$ as
\begin{equation}
\label{eq:eta}
\eta(\vek r,t)=\left\{
  \begin{array}{lll}
    0 & : & |v_{\rm D}|\le v_{cr}\\
    C\frac{\left(v_{\rm D}(\vek r,t)-v_{cr}\right)}{v_0}& : &
    |v_{\rm D}|> v_{cr}
  \end{array}
  \right.
\end{equation}

with the threshold $v_{cr}$ set to the higher, more realistic value
corresponding to increased resolution.
The authors applied
the model to the large-scale vertical current sheet in solar eruptive flares.

The simulations in Paper~I have been performed in dimensionless variables:
Spatial coordinates $x$, $y$, and $z$ has been expressed in units of
the current sheet half-width $L_{\rm A}$ at the photospheric level ($z=0$)
and the time has been normalised to the Alfv\'en transit time 
$\tau_{\rm A}=L_{\rm A}/V_{\rm A,0}$, where 
$V_{\rm A,0}=B_0/\sqrt{\mu_0 \rho_0}$ is the asymptotic value
($x\rightarrow\infty, z=0$) of the Alfv\'en speed at $t=0$. 
In order to relate the model to real solar conditions appropriate
scaling has been adopted in Paper~I. Based on the gravity-introduced
scale-height relation $L_{\rm A}=600$~km has been found. Due to this relation
both the dimension-less and SI units can be used further in our present paper.

In Paper~I the authors described how tearing and driven coalescence
instabilities lead to the fragmentation of an originally unstructured
flare current layer. As a consequence a
cascade of magnetic flux-ropes/plasmoids is formed from large to
consecutively smaller scales. The structuring of the current layer
leads to formation of multiple thin embedded current sheets. They can
host non-ideal regions where particles can be
accelerated. Enhanced numerical resolution revealed the
structure of dissipation regions: They form many thin channels of non-zero 
magnetic diffusivity. The spatial distribution 
and dynamics of these diffusivity channels has been further studied
via tracking of the X-points associated with the non-ideal regions. 
Analysis shows hierarchically-structured grouping of the dissipation 
channels and intermittency in their life-times and times for which
they are magnetically connected to the photosphere. 

The structure, distribution and dynamics of the non-ideal regions
should be reflected in specific observable features that we are going to
derive in the following sub-section.

\subsection{Derivation of model-specific observable features}
\label{subsect:modext}

Unfortunately, it is impossible to directly measure the magnetic field
or current density in the coronal current sheets (and at all not with the
resolution reached in the simulation done in Paper~I). Hence, we need some
indirect, more subtle comparison between observed and modelled
quantities. A possible way which may lead to indirect indication of 
current-sheet fragmentation has recently been presented by 
\citet{Nishizuka+:2009}. These authors study the structure of emission 
in flare ribbons, namely the distribution and dynamics of embedded
bright kernels. They conjecture possible relation of the resulting
power-law distributions found in their statistical analysis of
ribbon-kernel properties with the concept of 
\textit{fractal current sheet} \citep{Shibata+Tanuma:2001}.

We follow this idea from the other end. We start with the structure, 
distribution, and dynamics of diffusion/acceleration regions that are
specific to the cascading-reconnection model as they has been found in
Paper~I. Then, we relate these features to the structure and dynamics of
emission in flare ribbons. We proceed as follows:
As shown in Paper~I (Fig.~6), the reconnection X-points are associated
with thin channels of magnetic diffusivity. Hence, we take the X-points
as geometric representatives of dissipation regions. In these
regions electrons can be accelerated, e.g., by the DC electric fields.
However, also different acceleration mechanisms has been proposed --
e.g. \citet{Drake+:2005} suggest Fermi-type acceleration in
contracting plasmoids. We can include this type of acceleration into
account by following consideration: Extrapolating the results of
numerical simulations of reconnection cascade to smaller scales one
can imagine, that each dissipative region surrounding studied X-point 
in fact contains many unresolved
small-scale magnetic islands. Electron acceleration in these non-ideal
regions is then performed \citep[in line with ideas by][]{Drake+:2005}
via coalescence and shrinkage of these micro-plasmoids. Anyway,
electrons accelerated by Fermi-type mechanism inside the plasmoids are
trapped and can be released to the open field (which connects CS to
chromosphere) again in the vicinity of
the X-points, where they become demagnetized. For further drelated discussion
on electron acceleration in magnetic islands see, e.g. \citet{Oka+:2010}.

In our study we focus on electrons as we are interested in optical and UV/EUV
chromospheric response ascribed to electron beams.
These are then transported along the magnetic field lines either up-ward to
the solar corona or downward, until they reach dense layers of the
solar atmosphere, where their energy is thermalized. 
Here we concentrate on the down-ward transported electrons. 
In order to study the positions and dynamics of the points, where
these electrons reach the chromosphere, i.e. the expected bright kernels, 
we map found X-points to the bottom boundary of the simulation box using 
magnetic field. The magnetic field lines that go through given X-point
represent magnetic separatrices in our 2D model. Therefore, we are
seeking for intersections of magnetic separatrices with the bottom boundary.
In the following we will refer to this footpoints at the bottom boundary as to 
\textit{kernels}. Thus, from known position of all X-points (Fig.~7 in Paper~I)
and knowing the magnetic field we can calculate the positions $x_k(t)$ of all
kernels $k$ for each recorded time step $t$ of the simulation. Note, that
due to the 2D geometry of the model the foot-point/kernel position is
uniquely given by its $x$-coordinate (see Paper~I for details).


\begin{figure*}[t]
\epsscale{1.1}
\plotone{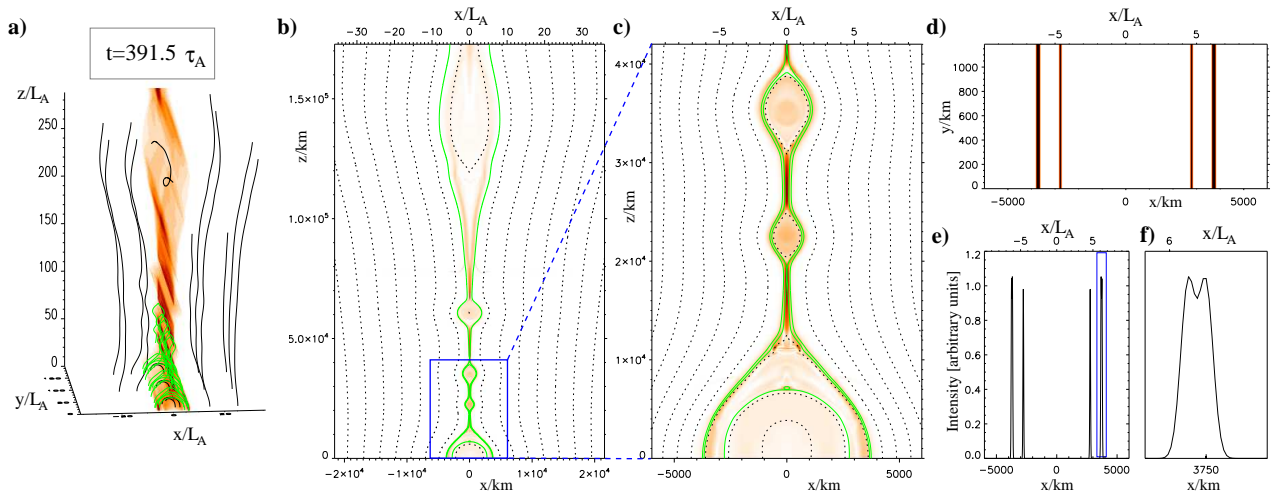}
\caption{ Model-specific consequences of cascading magnetic reconnection.
  (a) Global 3D magnetic and current-density structures 
  at $t=391.5\tau_{\rm A}$. The magnetic field (black lines),
  the separatrices that map the diffusive regions to the bottom boundary
  (green lines), and the current density (red colour scale) are shown. (b)
  Projection of (a) into the $xz$-plane. The $x$-axis shows positions both
  in the units of $L_{\rm A}$ (top) and in kilometres according to
  the scaling adopted in Section~\ref{subsect:modmhd}. (c) Enlarged view
  of the selected rectangle reveals double-structure of the outer pair
  of separatrices hitting the bottom boundary near $x\pm3750$~km. (d)
  Modelled view at two pairs of flare emission ribbons (inverted colour 
  scale, darker means higher intensity here). (e) Modelled
  emission profiles across the ribbons. (f) Detailed view of the
  outer-ribbon profile reveals its internal double-peaked structure.}
\label{fig:ribstruct}
\end{figure*}


The calculations of the chromospheric emission in a certain spectral line 
(e.g. H$_\alpha$ or C~{\sc iv}) in response to the bombardment of
electrons accelerated in a distant reconnection region is a difficult
task \citep{Kasparova+:2009, Varady+:2010}. In order to obtain an at
least qualitative output that could be later compared with
observations we use for ribbon-intensity distribution
the expression 
\begin{eqnarray}
\label{eq:halpha}
\nonumber
I_k(x,y,t)=I_0 \exp\left(-\frac{\left(x-x_k(t)\right)^2}{\Delta^2}\right)\\
I(x,y,t)=\sum_{k} I_k(x,y,t)\ ,
\end{eqnarray}
instead of description of all the complicated processes of electron
transport, energy deposition, chromospheric response, and radiative transfer.
The intensity $I$ is summed over all kernels $k$; $y$ is the second
(however invariant in our model) coordinate in the photosphere. 
The kernel size is set to $\Delta=0.1 L_{\rm A}$. Since the model is
capable neither to estimate the electron-beam flux nor the energy
thermalized in the kernel we set the intensity scale to $I_0=1$. We
will return to this point later in the discussion.

Let us now apply this procedures to the results of simulations
presented in Paper~I.


\section{Relation to observations}
\label{sect:results}

Fig.~\ref{fig:ribstruct} depicts the above-described procedure of
mapping of X-points to the photosphere along magnetic separatrices and
flare-ribbon emission calculation. Obtained results are shown for 
$t=391.5\tau_{\rm A}$. 
Panels (a) -- (c) show how three dissipative regions/X-points
embedded in the fragmented current layer are magnetically connected
to the bottom boundary and mapped by separatrices to the positions 
$x\approx\pm 2800$~km (one kernel) and $x\approx\pm 3750$~km (two
mutually close kernels).


\begin{figure}[t]
\epsscale{0.9}
\plotone{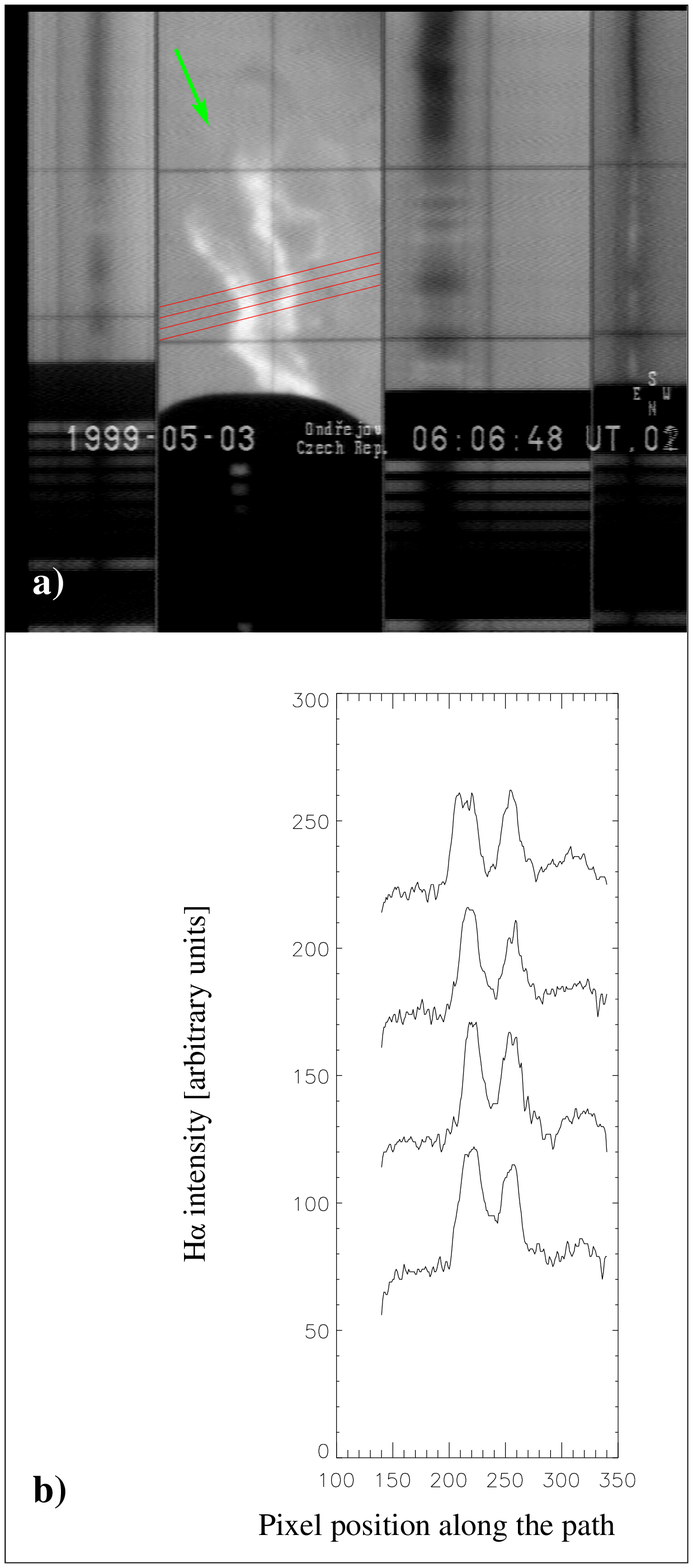}
\caption{Observation of flare H$_\alpha$ emission ribbons (a) and
  profiles of  H$_\alpha$ intensity (b) along selected paths (red lines in
  the panel (a)). The ribbon emission is structured (double-peaked?)
  -- c.f. Fig.~\ref{fig:ribstruct}. The green arrow points to the
  filament-like dark structure which projects to the space between
  flare ribbons. }
\label{fig:observ}
\end{figure}


A view at the flare ribbons obtained by applying 
the simple model described by Eq.~(\ref{eq:halpha}) is presented in 
Fig.~\ref{fig:ribstruct}(d). As the figure indicates, in some cases one
can observe even multiple pairs of flare ribbons. This is close to the idea of
chromospheric re-brightening studied by \citet{Miklenic+:2010}.
The inner separatrix
associated with the internal pair of ribbons is connected to the
X-point which appeared temporarily in the transversal current sheet
formed between merging plasmoid and the loop arcade (see also
Fig.~4 in Paper I). Figs.~\ref{fig:ribstruct}(e) and~(f) show the
emission intensity profile along the $x$-axis calculated according to
the model in Eq.~(\ref{eq:halpha}). The enlarged detail shown in
Fig.~\ref{fig:ribstruct}(f) reveals an internal double-peak structure
of the outer ribbons. Note, that a similar structuring is seen in 
observed H$_\alpha$ ribbons (Fig.~\ref{fig:observ}).

\begin{figure}[t]
\plotone{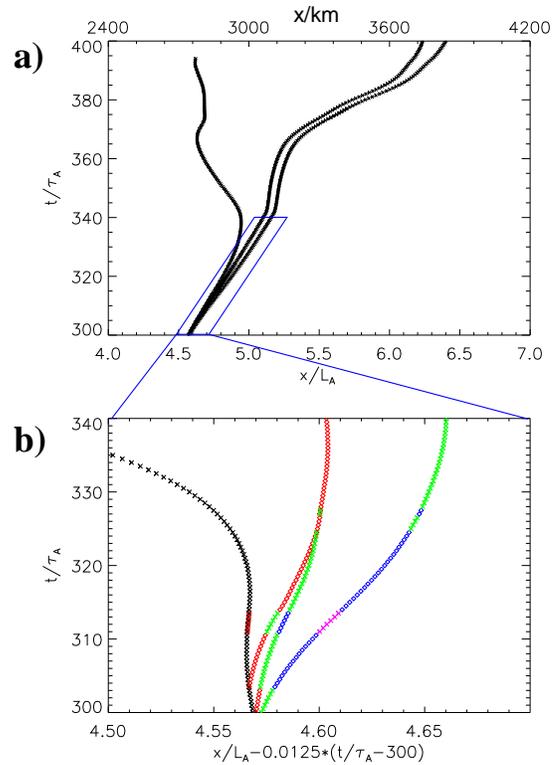}
\caption{Kinematics of flare ribbon kernels obtained by magnetic mapping
  of dissipation regions in cascading reconnection down to the
  photosphere. (a) View of all kernel positions in 
  t=$300\ \tau_{\rm A}$ ---$400\ \tau_{\rm A}$.
  Abscissa positions in kilometres (upper scale) and in units of $L_{\rm A}$. 
  (b) Detailed view of the area selected in (a) (the blue rhomboid). 
  The global motion of the group of kernels has been subtracted in
  order to reach larger zoom. For each time instant each kernel is
  depicted by unique color in order to distinguish between close
  (almost overlapping) kernels/foot-poits.} 
\label{fig:fps}
\end{figure}

In order to study the dynamics of kernels and associated X-points
during the entire recorded interval of evolution 
($300 \tau_{\rm A}$ --- $400 \tau_{\rm A}$) we tracked the moving positions
of all X-points and the magnetically-associated (mapped) kernels. 
Kinematics of the kernels (foot-points of separatrices) is depicted in
Fig~\ref{fig:fps}. Panel \ref{fig:fps}(a) presents a global picture of kernel
dynamics over the entire recorded interval. Because of the CS symmetry
only the right ribbon
positions are displayed. The spatial coordinate $x$, the distance
from the polarity-inversion line (PIL), 
is limited to the interval where the kernels actually occur.
Fig~\ref{fig:fps}(b) shows a detailed view of the area selected in
panel (a). In order to reach higher zoom the global trend
(motion) of the group of kernels (i.e. the increasing ribbon
separation as the flare proceeds) has been subtracted by applying
the transformation
$\overline{x}/L_{\rm A}=x/L_{\rm A}-0.0125(t/\tau_{\rm A}-300)$. 
X-points that are separated only by tiny magnetic islands can map 
practically to the same kernel. They then cannot be distinguished even in
the zoomed display. Therefore, for each time we take a set of kernel
positions (its size varies with time) and display every kernel in 
different colour. Fig.~\ref{fig:fps} shows various life-times of the
kernels and other aspects of their
dynamics, like bifurcation/merging. Note, that the life-time of the
kernel is not determined solely by the life-time of the associated
distant X-point but also by the processes in the current layer that can
change the magnetic connectivity of diffusion regions embedded in the
current layer to the photosphere (Fig.~7 in Paper I).


We have further extracted selected characteristics of the modelled kernel
dynamics and compare them with the observations of \citet{Nishizuka+:2009}.
For this sake we processed our model results in the same manner as
those authors by binning photosphere along the $x$-axis. For the bin
size we chose $\Delta x=0.05 L_{\rm A}$. For each time we
integrated the intensity given by Eq.~(\ref{eq:halpha}) over each
bin. Thus we obtained a proxy for ``light-curves'' of all bins.
Fig.~\ref{fig:lcurves} shows three examples of such light-curves for
bins centred around positions $x=4.58 L_{\rm A}$ (2750~km), 
$x=4.93 L_{\rm A}$ (2950~km), and $x=5.73 L_{\rm A}$ (3435~km).
Narrow peaks coming from the short-living X-points are superimposed
over the longer living bright kernels 
\citep[c.f. Fig.~2(b) in][]{Nishizuka+:2009}.

As in \citet{Nishizuka+:2009}, we then determined the peaks at
each light-curve and recorded their time of occurrence and their maximum
intensity. We calculated the statistical distributions of the peak 
intensities and the time-intervals between two subsequent peaks. The results
of this analysis are presented in Fig.~\ref{fig:distrib}. Despite of the
poor statistics (for our 1D mesh we identified only 68 peaks)
we obtained spectral indices $s=-1.48$ for the peak-intensity 
(panel \ref{fig:distrib}(a)) and $s=-1.73$ for the time interval
between subsequent peaks (Fig.~\ref{fig:distrib}(b)).


\section{Discussion and conclusions}
\label{sect:conclusions}

The 'standard' CSHKP picture of solar eruptive flares involves the formation
of a global flare current sheet (CS) behind a ejected filament followed by
magnetic reconnection in this CS. Nevertheless,
deeper study of this scenario invokes further questions: 1)
How is the energy accumulated in flares in relatively large-scale
structures (global flare current layer) transferred to the
dissipative (kinetic) scales, 2) How can we observe regular dynamics
in eruption/flare on large scales, just in line with the CSHKP
scenario, and the signatures of fragmented energy release at the same moment, 
and 3) How can the large fluxes of accelerated electrons inferred
from HXR observations be reconciled with a single, relatively small
diffusion region assumed in classical CSHKP model.

The simulations presented in Paper~I showed, that a cascading reconnection
resulting from the formation and interaction of plasmoids/flux-ropes 
can address these three fundamental questions at once and that it
might provide viable scenario for magnetic energy dissipation in
large-scale systems, like solar (eruptive) flares. 

In the present paper we evaluated the relevance of this model for
actual solar flares. For this sake we derived observable model predictions
that are specific for the cascading-reconnection scenario, and
searched for the predicted features in observed data. Since
multiple acceleration regions embedded in the global flare current
layer are -- via magnetic field-line mapping -- inherently connected with 
the structured emission in the flare ribbons, detailed study of the
ribbon kernels might bring some evidence for the cascading processes
in the current layer above the flaring region. This idea is
illustrated by Fig.~\ref{fig:ribstruct} which shows the structuring of
modelled emission in ribbons. Fig.~\ref{fig:ribstruct} further
indicates, that the larger-scale 
plasmoid interacting with the flare-loop arcade can temporarily
form a second pair of ribbons. The plasmoid/loop-arcade
interaction may play significant role in flare dynamics. Another
indirect evidence of such interaction based on analysis of series of
X-ray images of the limb flare has been presented by
\citet{Kolomanski+Karlicky:2007} and \citet{Milligan+:2010}.

Similar structuring of H$_\alpha$ emission as presented in
Figs.~\ref{fig:ribstruct}(d)--(f) is, indeed, visible in
Fig.~\ref{fig:observ}. Yet another
interesting feature in Fig.~\ref{fig:observ} is the dark filament-like 
structure (indicated by an arrow). Since the original large-scale
filament whose eruption initiated the flare was already far away at the time of
observation, the absorption feature can be interpreted
as a manifestation of one of the secondary plasmoids formed in the
global current layer in line with the presented
cascading-fragmentation scenario. The secondary plasmoids
represent enhanced-density structures. Density increase in
connection with consequent faster radiative cooling might lead to the
detectable H$_\alpha$ absorption. The secondary plasmoid can be
consequently manifested as a darker feature at the background of
relatively brighter chromosphere.

A sophisticated study of the structure of flare ribbons was
presented by \cite{Nishizuka+:2009}. These authors studied the statistical
properties of emission kernels found inside the ribbons and they 
found a power-law
distribution of the studied kernel characteristics. A power-law
distribution can be a signature of self-organised criticality
evolution \citep{Aschwanden:2002, Vlahos:2007}. It can also indicate a
\textit{fractal current sheet} decay \citep{Shibata+Tanuma:2001}. 
The latter scenario is in fact favoured by the subsequent analysis of 
\citet{Nishizuka+:2010} which relates the energy release measured by
means of HXR flux and ejection of multiple plasmoids formed in the fragmenting
current sheet.

\begin{figure}[t]
\plotone{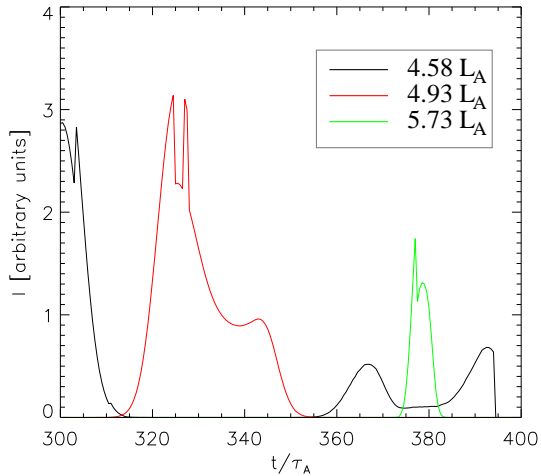}
\caption{Modelled light-curves for three selected bins at positions
  $x=4.58 L_{\rm A}$ (2750~km), $x=4.93 L_{\rm A}$ (2950~km), 
  and $x=5.73 L_{\rm A}$ (3435~km).}
\label{fig:lcurves}
\end{figure}

In our present paper we have established the connection between the
hierarchical distribution of diffusion regions formed by cascading
reconnection and the statistical properties of 
emission kernels from first principles. We have studied
the dynamics (positions, and the mechanisms and times of 
creation/annihilation) of O- and X-type
points in the current layer as well as of the kernel-points associated
with the diffusion regions (X-points) by means of magnetic field-line mapping. 
We then associated each found kernel with the emission according to 
Eq.~(\ref{eq:halpha}). We chose this simple model as we do not have
any information about the accelerated electron fluxes absorbed in the
kernel, nor can we easily calculate the chromospheric response in
terms of (e.g., H$_\alpha$) emission. Our choice of
the unified peak intensity $I_0=1$ for all kernels can be justified
following way: After all, the particle acceleration is performed by
kinetic processes at the small-scale end of the cascade. Suggested
mechanism \citep[see e.g.][]{Drake+:2005} involves, e.g. 
kinetic coalescence and shrinkage of small-scale magnetic 
islands/plasmoids. In the dissipative range of scales the slope of the
energy spectrum  (see Fig.~1 in Paper~I) is very steep. 
Hence, the size of plasmoids
at kinetic scale can be expected more or less uniform as indicated also
by PIC simulations. In such case the elementary injections of
accelerated particles are, perhaps, roughly of
the same magnitude. The intermittent, spiky structure of observed emission
profiles is then the result of overlapping of many of these elementary
injection peaks \citep{Aschwanden:2002}. Indeed, we observe this
feature in examples of the modelled light-curves obtained at  fixed points at
the bottom boundary (Fig.~\ref{fig:lcurves}). Despite of the intensity
produced by the elementary particle injection being normalised to $I_0=1$,
the overlapping elementary peaks produce spiky intensity profiles ranging
from $I=0$ to $I > 3$.  


\begin{figure}[t]
\plotone{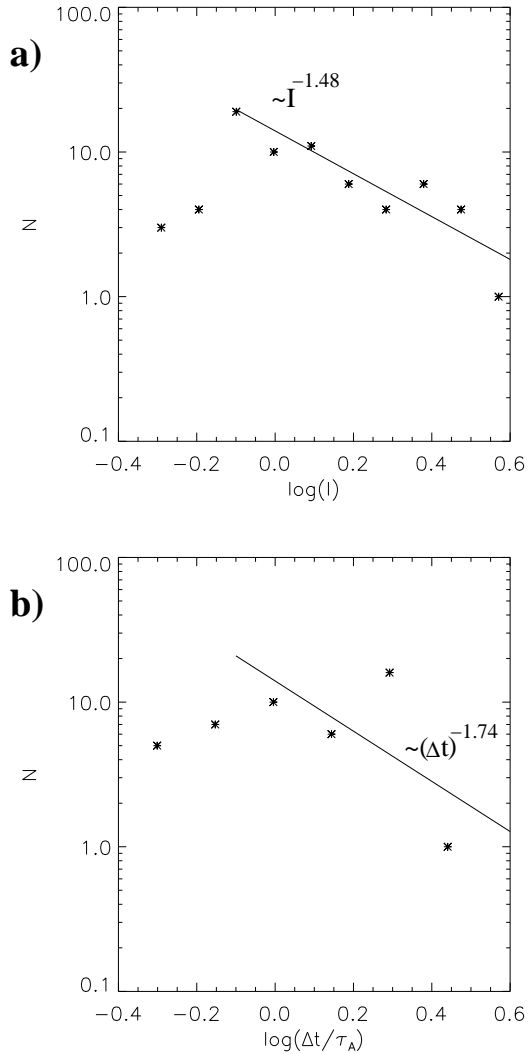}
\caption{Statistical distribution of the modelled intensity (a) and the
  time-interval between subsequent emission peaks (b) for accumulated
  light-curves from all mesh-boxes. $N$ indicates the number of peaks
  with the property within the bin.}
\label{fig:distrib}
\end{figure}


In order to directly compare our results with the findings 
of \citet{Nishizuka+:2009} we have performed the same statistical 
analysis of the peaks in our artificial light-curves. We have found
the statistical distributions of the emission peak 
intensities and of the time-lag between subsequent emission peaks
(Fig.~\ref{fig:distrib}). Because of the 2-D geometry of underlying
MHD model we have obtained 1-D distribution of emission kernels. 
Consequently we have found smaller number of the peaks. 
Despite of the resulting poor statistics we
could conclude, that the slopes of the power-law
distributions obtained by our model agree surprisingly well
with those found by \citet{Nishizuka+:2009} for observed
light-curves: For the statistical distribution of the peak intensities
we have found spectral index $s=-1.48$ (\citet{Nishizuka+:2009}
obtained $s=-1.5$) and 
for the time-lag between peaks we obtained $s=-1.74$ 
($s=-1.8$ in \citet{Nishizuka+:2009}). We are aware that our
statistics is, however, rather poor as it operates with only 68 peaks 
found in total. Thus, it would be premature to take this agreement as 
conclusive. Nevertheless the result indicates that the statistical 
properties of emission kernels resulting from cascading reconnection
and of those observed are -- at least -- not in direct contradiction. 

Results of presented comparison between the modelled features
attributed specifically to cascading reconnection and their observed
counterparts supports -- in our view -- the relevance of
cascading reconnection processes studied in Paper~I for reconnection
in solar atmosphere. 
Agreements found between the statistical properties of modelled and
observed emission kernels invokes, however, further questions. 
The underlying 2.5D MHD model ignores 
structuring of the current sheet in the $y$-direction (along the
PIL). This result might indicate that modes with $k_y>0$ (in the
off-plane direction) may not play too
significant role in current-layer fragmentation in solar flares, 
or at least, that they do
not reach the smallest scales to substantially influence the
dynamics (life-time) of kinetic-scale diffusion regions. This issue
can be addressed, however, only within the framework of a full 3D MHD model.


\acknowledgments 
This research was performed under the support of the
European Commission through the SOLAIRE Network (MTRN-CT-2006-035484)
and the grants P209/10/1680, 205/09/1469, 205/09/1705, P209/10/1706 of
the Grant Agency of the Czech Republic, by the grant 300030701 of the 
Grant Agency of the Czech Academy of Science, and the research project 
AV0Z10030501 of Astronomical Institute of the Czech Academy of Science.
    
The authors thanks to anonymous referee, whose comments where very
helpful in improving quality of the paper.



\begin{thebibliography}{25}
\expandafter\ifx\csname natexlab\endcsname\relax\def\natexlab#1{#1}\fi

\bibitem[{{Aschwanden}(2002)}]{Aschwanden:2002}
{Aschwanden}, M.~J. 2002, \ssr, 101, 1

\bibitem[{{B{\'a}rta} {et~al.}(2010){B{\'a}rta}, {B{\"u}chner}, {Karlick{\'y}},
  \& {Sk{\'a}la}}]{Barta+:2010b}
{B{\'a}rta}, M., {B{\"u}chner}, J., {Karlick{\'y}}, M., \& {Sk{\'a}la}, J.
  2010, \apj, submitted, 1011.4035

\bibitem[{{Bhattacharjee} {et~al.}(2009){Bhattacharjee}, {Huang}, {Yang}, \&
  {Rogers}}]{Bhattacharjee+:2009}
{Bhattacharjee}, A., {Huang}, Y., {Yang}, H., \& {Rogers}, B. 2009, Physics of
  Plasmas, 16, 112102, 0906.5599

\bibitem[{{B{\"u}chner}(2006)}]{Buchner:2006}
{B{\"u}chner}, J. 2006, Space Science Reviews, 124, 345

\bibitem[{{Drake} {et~al.}(2005){Drake}, {Shay}, {Thongthai}, \&
  {Swisdak}}]{Drake+:2005}
{Drake}, J.~F., {Shay}, M.~A., {Thongthai}, W., \& {Swisdak}, M. 2005, Physical
  Review Letters, 94, 095001.1

\bibitem[{{Huang} \& {Bhattacharjee}(2010)}]{Huang+Bhattacharjee:2010}
{Huang}, Y., \& {Bhattacharjee}, A. 2010, Physics of Plasmas, 17, 062104,
  1003.5951

\bibitem[{{Huang} {et~al.}(2010){Huang}, {Bhattacharjee}, \&
  {Sullivan}}]{Huang+:2010}
{Huang}, Y.-M., {Bhattacharjee}, A., \& {Sullivan}, B.~P. 2010, ArXiv e-prints,
  1010.5284

\bibitem[{{Karlick{\'y}} {et~al.}(2010){Karlick{\'y}}, {B{\'a}rta}, \&
  {Ryb{\'a}k}}]{Karlicky+:2010}
{Karlick{\'y}}, M., {B{\'a}rta}, M., \& {Ryb{\'a}k}, J. 2010, \aap, 514, A28+

\bibitem[{{Karlick{\'y}} {et~al.}(2000){Karlick{\'y}}, {Ji{\v r}i{\v c}ka}, \&
  {Sobotka}}]{Karlicky+:2000}
{Karlick{\'y}}, M., {Ji{\v r}i{\v c}ka}, K., \& {Sobotka}, M. 2000, \solphys,
  195, 165

\bibitem[{{Ka{\v s}parov{\'a}} {et~al.}(2009){Ka{\v s}parov{\'a}}, {Varady},
  {Heinzel}, {Karlick{\'y}}, \& {Moravec}}]{Kasparova+:2009}
{Ka{\v s}parov{\'a}}, J., {Varady}, M., {Heinzel}, P., {Karlick{\'y}}, M., \&
  {Moravec}, Z. 2009, \aap, 499, 923, 0904.2084

\bibitem[{{Ko{\l}oma{\'n}ski} \&
  {Karlick{\'y}}(2007)}]{Kolomanski+Karlicky:2007}
{Ko{\l}oma{\'n}ski}, S., \& {Karlick{\'y}}, M. 2007, \aap, 475, 685

\bibitem[{{Lin} \& {Forbes}(2000)}]{Lin+Forbes:2000}
{Lin}, J., \& {Forbes}, T.~G. 2000, \jgr, 105, 2375

\bibitem[{{Loureiro} {et~al.}(2007){Loureiro}, {Schekochihin}, \&
  {Cowley}}]{Loureiro+:2007}
{Loureiro}, N.~F., {Schekochihin}, A.~A., \& {Cowley}, S.~C. 2007, Physics of
  Plasmas, 14, 100703, arXiv:astro-ph/0703631

\bibitem[{{Magara} {et~al.}(1996){Magara}, {Mineshige}, {Yokoyama}, \&
  {Shibata}}]{Magara+:1996}
{Magara}, T., {Mineshige}, S., {Yokoyama}, T., \& {Shibata}, K. 1996, \apj,
  466, 1054

\bibitem[{{Miklenic} {et~al.}(2010){Miklenic}, {Veronig}, {Vr{\v s}nak}, \&
  {B{\'a}rta}}]{Miklenic+:2010}
{Miklenic}, C.~H., {Veronig}, A.~M., {Vr{\v s}nak}, B., \& {B{\'a}rta}, M.
  2010, \apj, 719, 1750

\bibitem[{{Milligan} {et~al.}(2010){Milligan}, {McAteer}, {Dennis}, \&
  {Young}}]{Milligan+:2010}
{Milligan}, R.~O., {McAteer}, R.~T.~J., {Dennis}, B.~R., \& {Young}, C.~A.
  2010, \apj, 713, 1292, 1003.0665

\bibitem[{{Nishizuka} {et~al.}(2009){Nishizuka}, {Asai}, {Takasaki},
  {Kurokawa}, \& {Shibata}}]{Nishizuka+:2009}
{Nishizuka}, N., {Asai}, A., {Takasaki}, H., {Kurokawa}, H., \& {Shibata}, K.
  2009, \apjl, 694, L74

\bibitem[{{Nishizuka} {et~al.}(2010){Nishizuka}, {Takasaki}, {Asai}, \&
  {Shibata}}]{Nishizuka+:2010}
{Nishizuka}, N., {Takasaki}, H., {Asai}, A., \& {Shibata}, K. 2010, \apj, 711,
  1062

\bibitem[{{Oka} {et~al.}(2010){Oka}, {Phan}, {Krucker}, {Fujimoto}, \&
  {Shinohara}}]{Oka+:2010}
{Oka}, M., {Phan}, T., {Krucker}, S., {Fujimoto}, M., \& {Shinohara}, I. 2010,
  \apj, 714, 915, 1004.1154

\bibitem[{{Samtaney} {et~al.}(2009){Samtaney}, {Loureiro}, {Uzdensky},
  {Schekochihin}, \& {Cowley}}]{Samtaney+:2009}
{Samtaney}, R., {Loureiro}, N.~F., {Uzdensky}, D.~A., {Schekochihin}, A.~A., \&
  {Cowley}, S.~C. 2009, Physical Review Letters, 103, 105004, 0903.0542

\bibitem[{{Shepherd} \& {Cassak}(2010)}]{Shepherd+Cassak:2010}
{Shepherd}, L.~S., \& {Cassak}, P.~A. 2010, Physical Review Letters, 105,
  015004, 1006.1883

\bibitem[{{Shibata} \& {Tanuma}(2001)}]{Shibata+Tanuma:2001}
{Shibata}, K., \& {Tanuma}, S. 2001, Earth, Planets, and Space, 53, 473,
  arXiv:astro-ph/0101008

\bibitem[{{Uzdensky} {et~al.}(2010){Uzdensky}, {Loureiro}, \&
  {Schekochihin}}]{Uzdensky+:2010}
{Uzdensky}, D.~A., {Loureiro}, N.~F., \& {Schekochihin}, A.~A. 2010, Physical
  Review Letters, 105, 235002, 1008.3330

\bibitem[{{Varady} {et~al.}(2010){Varady}, {Ka\v{s}parov\'a}, {Moravec},
  {Heinzel}, \& {Karlick\'y}}]{Varady+:2010}
{Varady}, M., {Ka\v{s}parov\'a}, J., {Moravec}, Z., {Heinzel}, P., \&
  {Karlick\'y}, M. 2010, IEEE Transactions on Plasma Science, 38, 2249

\bibitem[{{Vlahos}(2007)}]{Vlahos:2007}
{Vlahos}, L. 2007, in Lecture Notes in Physics, Berlin Springer Verlag, Vol.
  725, Lecture Notes in Physics, Berlin Springer Verlag, ed. K.-L. {Klein} \&
  A.~L. {MacKinnon}, 15--31

\end{thebibliography}

\end{document}